\documentclass[pdflatex,sn-mathphys]{sn-jnl}

\jyear{2021}%

\theoremstyle{thmstyleone}%
\theoremstyle{thmstyletwo}%

\theoremstyle{thmstylethree}%

\begin{document}

\title[Monu et al.]{Detecting Tropical Cyclone from the basic overview of life cycle of Extremely Severe Cyclonic Storm, Tauktae}


\author[1]{\fnm{Monu} \sur{Yadav}} \email{yadavm012@gmail.com}

\author*[1]{\fnm{Laxminarayan} \sur{Das}}\email{lndas@dce.ac.in}


\affil*[1]{\orgdiv{Department of Applied Mathematics}, \orgname{Delhi Technological University}, \orgaddress{ \state{New Delhi}, \country{India}}}




\abstract{The Extremely Severe Cyclonic Storm (ESCS) Tauktae, which made landfall on the Gujarat coast on May 17, 2021, is discussed in the current study. The analysis is based on INSAT-3D and passive microwave (PMW) images, focusing on the cyclone's eye characteristics and intensity. The satellite images and products are utilized to determine the cyclone's intensity and the specific characteristics of its eye.
	
The study showcases the variations in intensity and eye features of the cyclone throughout its life span, providing insights into the process of tropical cyclone intensification. The paper is structured to cover the methods for analysis, an overview of ESCS Tauktae's life cycle, the regulation of intensity based on Dvorak's technique, intensity estimation using ADT9.0 and comparison with SATCON method and Indian Meteorological Department (IMD) provided best track data. Furthermore, the formation of Tauktae's eye, timeframes for eye scenes, comparison with sea surface temperature data, and concluding thoughts are also discussed.}




\maketitle

\section{Introduction}\label{sec1}
Tropical cyclones (TCs) not only bring violent weather but also wreak havoc along coastal regions, causing storm surges, flooding, strong winds, heavy rainfall, thunderstorms, and lightning. These destructive conditions lead to significant economic losses and loss of life worldwide each year. Accurate prediction of TC formation is crucial to mitigate these losses and enable effective preventive measures. The term ``Cyclone" is commonly used to refer to a cyclonic system with winds equal to or exceeding 34 knots (62 km/h), as defined by the World Meteorological Organization (WMO) \cite{bib32}. Tropical storms are defined based on maximum sustained wind speed (MSW), although the naming conventions may differ across regions.

In the East Arabian Sea and adjacent Lakshadweep area, an extremely strong cyclonic storm called ESCS, TAUKTAE formed, making landfall near latitude 20.8°N and longitude 71.1°E, northeast of Diu with wind speeds of 160-170 km/h and gusts up to 185 km/h. ESCS, TAUKTAE rapidly intensified from 65 knots on May 16th to 100 knots on May 17th over the sea. It significantly affected marine life and activities. In the northern Indian Ocean, it was the first cyclonic storm of 2021, and since 1961, it was the most intense cyclone recorded in the satellite era, surpassing the intensity of the Kandla cyclone in 1998.

Detecting and predicting the maximum intensity of TCs over the ocean is crucial for providing accurate early warnings and enabling prompt responses, especially considering the smaller size of the basin and the socioeconomic vulnerability of the region. Satellites, in conjunction with computing devices, play a crucial role in detecting and predicting the intensity of these weather events. However, numerical weather prediction models \cite{bib22,bib23}, dynamical-statistical models \cite{bib24}, and operational forecasts \cite{bib25,bib26} still face challenges in accurately predicting TC intensity, particularly rapid intensification \cite{bib27}.

This study aims to identify various technical characteristics of ESCS by utilizing satellite data from INSAT-3D and microwave images. Focusing on the eye pattern of the TC, the primary objective is to examine the relationship between TC, Tauktae eye characteristics and storm intensity. Additionally, different methods were employed to calculate TC intensity and compare them with the best track results. Only a few studies have explored the relationship between specific eye features and TC intensification in the Arabian Sea. The Dvorak method \cite{bib4,bib5} utilizes satellite imagery to estimate TC intensity based on the position and associated characteristics of the eye. Although there is no standardized measure to gauge the well-formedness or raggedness of TC eyes, generally, a sharper eye indicates higher TC intensity.

Geometric characteristics of TC eyes have garnered increasing attention. Satellite and radar observations distinguish TC eyes from one another, revealing their circular or elliptical shapes \cite{bib15,bib27,bib28,bib29,bib30}. Furthermore, large-eyed TCs exhibit a slower rate of eyewall contraction when their intensification is preceded by a reduction in eye area \cite{bib15}.

The geometric characteristics of the eye, as described above, are crucial components of the inner core of TCs. These characteristics can be examined using INSAT-3D and PMW images alongside TC intensity measurements for ESCS, Tauktae.

Section \ref{sec2} provides an overview of the data and methodology used in analyzing the cyclone, including a brief description of the Dvorak Method. Section \ref{sec3} presents the result and discussions, while Section \ref{sec4} concludes the study.
\section{Data And Methods} \label{sec2}
Data for this study was primarily collected from Indian Geostationary Satellite Images and Passive Microwave (PMW) Images. Hourly data from INSAT-3D and INSAT-3DR satellites, synoptic hourly data from bulletins and reports provided by Regional Specialized Meteorological Centre (RSMC), New Delhi \cite{bib1}, microwave images from the NRL TC website \cite{bib2}, and satellite bulletins \cite{bib3} were used to analyze various characteristics of ESCS, Tauktae.

The analysis process involved using Meteorological Image Analysis Software (MS) and Dvorak's Technique \cite{bib4,bib5} to examine satellite images. MS, developed at the Space Application Center, was used to assess eye characteristics such as eye temperature, eye diameter, and the semi-major and semi-minor axes of the eye to determine the Eye Roundness Value (ERV) of ESCS, Tauktae.

According to Dvorak's intensity scale (T number), different phases of tropical disturbances are categorized as shown in Table \ref{tab1}. The intensity of ESCS, Tauktae was analyzed based on T numbers derived from INSAT-3D/3DR visible and IR images on a three-hourly basis \cite{bib4,bib5}. Additionally, the T number estimated from microwave images was examined to improve the accuracy of T number estimation by better determining the TC's center location.

The intensity of ESCS, Tauktae was also estimated using the Advanced Dvorak Technique (ADT) \cite{bib16}. ADT (9.0) accurately represents the situation during rapid intensification and weakening. The ADT and Manual T Number results were analyzed and compared with the best track intensity provided by RSMC, New Delhi. ADT values from CIMSS (Cooperative Institute of Meteorological Satellite System) \cite{bib17} were collected for ESCS, Tauktae.

The intensity of ESCS, Tauktae was also estimated using SATCON method \cite{bib31}, which is an approach for estimating cyclone intensity. The SATCON method result was compared with the best track intensity from the ESCS, Tauktae study conducted by RSMC, New Delhi.

Dvorak's method \cite{bib4} of estimating TC intensity is effectively applied by observing the eye in satellite imagery. The study focuses on analyzing various aspects of the eye using INSAT-3D and PMW images and their relationship to the intensification and weakening of ESCS.

Eye size is an important structural characteristic of a TC, and its effect on ESCS, Tauktae is examined. Strengthened eyewall convection as a TC intensifies leads to increased heating in the eye, resulting in a decrease in central pressure \cite{bib18,bib19,bib20}. The resulting contraction of the pressure gradient around the maximum wind causes the eye and eyewall to contract \cite{bib21}. The study examines the size, shape, and roundness of the eye, where the eye roundness value (ERV) is calculated as the
\begin{equation*}
	ERV = \sqrt{(a^2  - b^2)/a^2}      ; a,b \in R^+
\end{equation*} with a and b representing the semi-major and semi-minor axes of the fitted ellipse, respectively. Eyes frequently occur in the trajectory of the cyclonic storm \cite{bib15}.

To trace ERVs, this study utilizes INSAT-3D and PMW images instead of radar images, which are typically used. The accuracy of ERV determination from satellite imagery has improved due to higher-resolution images and DIGI (Digital Graphic User Interface) platforms. PMW images with spatial resolutions of 12.5-15 km and INSAT-3

\begin{table}[ht]
    \centering
    \begin{tabular}{|c|c|c|}
    \hline
     Phases& T Number& Maximum Wind Speed  \\
     \hline
     Low pressure area (WML)& T1.0 & $<$ 17 kts   \\
     \hline
     Depression (D) & T1.5 & 17-27 kts  \\
     \hline
     Deep Depression (DD) & T2.0 & 28-33 kts  \\
     \hline
     Cyclonic Storm (CS) & T2.5-T3.0 & 34-47 kts  \\
     \hline
     Severe CS (SCS) & T3.5 & 48-63 kts  \\
     \hline
     Very Severe CS (VSCS) & T4.0-T4.5 & 64-89 kts  \\
     \hline
     Extremely Severe CS (ESCS) & T5.0-T6.0 & 90-119 kts  \\
     \hline
     Super CS & T6.5 -T8.0 & $>$ 120 kts  \\
     \hline
\end{tabular}
\vspace{2mm}
    \caption{Different phases of tropical disturbances clasifed by Indian Meteorological Department}
    \label{tab1}
\end{table}

\section{Results and Discussion} \label{sec3}
\subsection{Overview of the life of ESCS, Tauktae}
ESCS, Tauktae originated from the remnants of a Low-Level circulation over the Arabian Sea on May 13th, which was designated as ``D" at 1200 UTC by RSMC, New Delhi (2021). It underwent intensification and became a ``DD" at 1200 UTC on May 14th. Moving in a northeast direction with increasing intensity, it reached ``CS" status at 1800 UTC on May 14th and was subsequently named ``Tauktae". On May 15th, Tauktae further intensified and attained Very Severe Cyclonic Storm (VSCS) status later in the day. By the following morning, it had become a VSCS and continued to move predominantly northward. Rapid intensification was observed on the morning of May 16th. Early on May 17th, it further intensified into an Extremely Severe Cyclonic Storm (ESCS) and reached its peak intensity. As it progressed towards the Gujarat coast, Tauktae weakened, followed by a restrengthening phase before making landfall. After landfall, as the storm moved northeastward, Tauktae gradually weakened as it moved inland. By May 19th, it had weakened to a Well-Marked Low (WML) over the northeast region.

During the satellite era (1961-2021), Cyclone Tauktae was the most intense cyclone after the Kandla storm in 1998. At landfall, Cyclone Tauktae had a similar intensity to the June 1998 Kandla cyclone, with sustained winds of approximately 160-170 kmph and gusts of up to 185 kmph. Although Tauktae reached a higher lifetime maximum intensity, with gusts reaching 210 kmph in the early morning to afternoon of May 17th, it experienced gusts of 180-190 kmph during the early morning to midafternoon period.

This rare cyclone, moving parallel to the west coast and crossing Gujarat, resulted in adverse weather conditions and caused significant damage along the west coast states of the country.

\begin{figure}
    \centering
    \includegraphics[width = 0.8\textwidth]{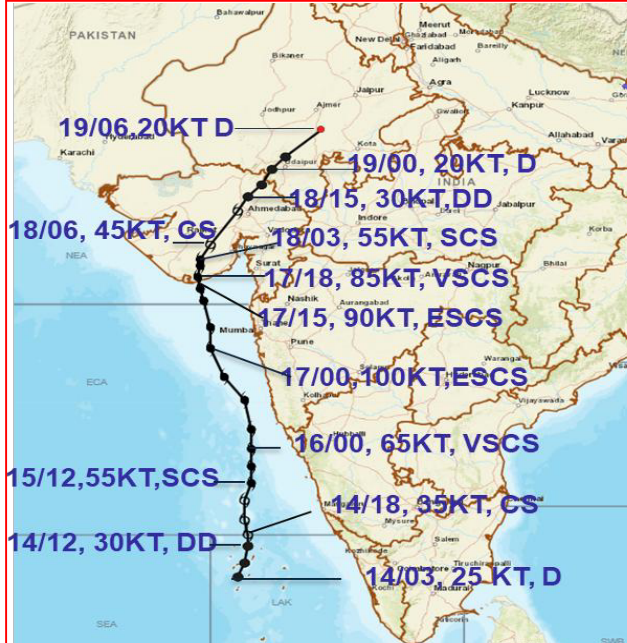}
    \caption{Observed track of ESCS, ``Tauktae" }
    \label{fig2}
\end{figure}

Tauktae traveled a distance of approximately 1880 km during its lifetime. Please refer to Figure \ref{fig2} for the observed tracking of ESCS, Tauktae.
\subsection{Intensity variations for Tauktae according to Dvorak's technique}
In this study, we utilized visible/IR images from geostationary satellites (INSAT-3D \& 3DR) along with the findings of \cite{bib5} to assess the intensity of ESCS, Tauktae. The initial pattern observed in Tauktae's formation was the ``Shear Pattern", which later transitioned into the ``Eye pattern" and persisted until landfall.

At 1300 UTC on May 15th, ESCS, Tauktae exhibited an ``Eye Pattern" with a corresponding intensity of T3.5, in agreement with the findings of \cite{bib9}. The formation of the CDO (Central Dense Overcast) pattern preceded the development of the eye, supporting previous observations that tropical cyclones tend to form an eye within 48 hours of attaining tropical strength. The eye structure was first observed in the visible imagery of INSAT-3D on May 16th at 0600 UTC, approximately 33 hours after reaching tropical storm strength on May 14th. The eye temperature was measured at -13 degrees Celsius at 0900 UTC on May 16th, contributing to a further increase in intensity to T4.5. Subsequently, as the eye sharpened and continued to warm, the intensity reached T5.0 at 2100 UTC on May 16th, classifying it as an ``ESCS" category cyclone. According to \cite{bib5}, a final T.No. of T5.5 was estimated.

By employing microwave images to determine the center and eye characteristics, the intensity was computed and found to align with the results obtained using the Dvorak technique \cite{bib5} on IR images. The microwave images obtained from NAVY NRL and CIRA web pages provided clear information about the eye features and the center of the tropical cyclone. Throughout the lifecycle of ESCS, Tauktae, the intensity calculation based on the accurate center determined through microwave images was consistent with the application of the Dvorak technique \cite{bib4, bib5}.
\subsection{Based on ADT9.0, the intensity of TC, Tauktae}
\begin{figure}
    \centering
    \includegraphics[width = 0.9\textwidth]{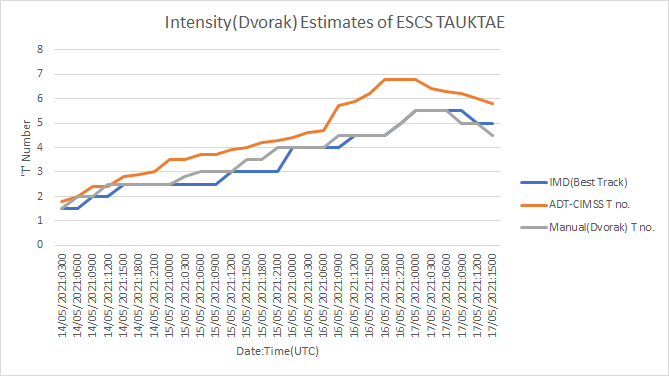}
    \caption{IMD Best Track and ADT (CIMSS) T numbers, and Manual (Dvorak) T numbers estimate intensity of ESCS-Tauktae}
    \label{fig1}
\end{figure}

The ADT (9.0) intensity exhibits an overestimation of up to T3.0, as shown in Figure \ref{fig1}. When compared to data provided by IMD and the Manual Dvorak T. No., there is a discrepancy of approximately T1.0, consistent with the findings of \cite{bib10}. Their research also indicated that ADT (Version 8.2.1) tends to overestimate intensity by up to T3.0 or T2.5. By transitioning from the original ``Curve Band" pattern to an ``Eye" pattern, ADT effectively captures the Best Track Intensity.

However, during the weakening phase, ADT intensity estimates slightly underestimated the IMD best track by approximately 0.5 T. Number. Moreover, in the preliminary development phase, ADT's Maximum Sustained Wind (MSW) was overestimated, whereas in the weakening phase, it was underestimated compared to the best track MSW. Specifically, during the initial development phase, the best track exhibited an overestimation of approximately 10-30 knots, while during the weakening phase, it displayed an underestimation of about 5-30 knots.

\subsection{On the basis of SATCON method, the intensity of TC, Tauktae}
\begin{figure}
    \centering
    \includegraphics[width = 0.9\textwidth]{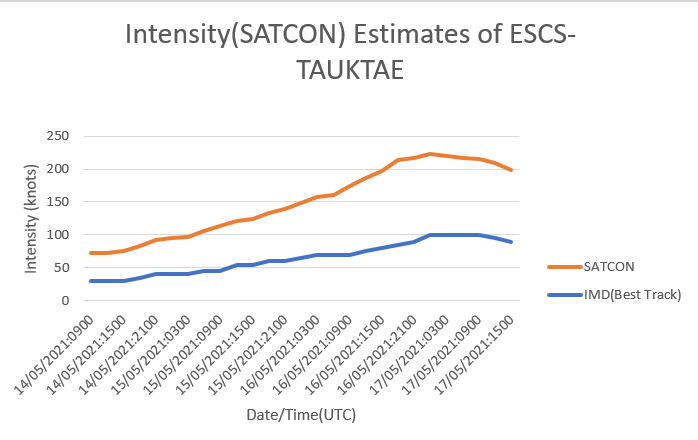}
    \caption{Intensity estimate by SATCON method and IMD of ESCS-Tauktae}
    \label{fig3}
\end{figure}

Figure \ref{fig3} depicts the overestimation of SATCON method based intensity. A comparison with the IMD best track data reveals that TC, Tauktae was consistently overestimated at both the onset and conclusion of its lifecycle. The overestimation tends to be more pronounced during the initial stage, reaching its peak at around 30 knots before gradually decreasing to around 10 knots.
\subsection{Description of ESCS, Tauktae's eye}

\subsubsection{Description of the formation of an `eye'}
The eye of Tropical Cyclone Tauktae was initially spotted in a microwave image taken by MSG-1 SEVIRI at 1300 UTC on May 15th, 2021, approximately 17 hours prior to its appearance in the INSAT-3D image. This observation is consistent with previous research findings \cite{bib6, bib7}. Microwave image scans were utilized to assess the cyclone's duration and detect the presence of cirrus clouds obstructing the eye. At 0900 UTC on May 15th, the Estimated Central Pressure (ECP) was recorded as 990 hPa, along with a Maximum Sustained Wind (MSW) of 45 knots and a 10 hPa pressure drop.

Moreover, analysis of ECP data from the Atlantic Ocean reveals that new eye formations typically occur within the range of 987 to 997 hPa, supporting the findings presented in \cite{bib6}. Additionally, the characteristics of Tauktae's eye align with the minimum central pressure reported in \cite{bib8} for the South China Sea region, ranging from 950 to 990 hPa. The initial measurement of Tauktae's eye temperature, taken from the microwave image captured by MSG-1 SEVIRI on May 15th at 1300 UTC, was recorded as -12.8 degrees Celsius.
\subsubsection{How the eyes are structured}
An analysis of the eye structure was conducted using satellite imagery from INSAT-3D and the Microwave satellite (PMW). The initial observation of the eye in the INSAT-3D satellite was recorded at 0600 UTC on May 16th, while in the Microwave satellite, it was observed on May 15th at 1300 UTC. By 2100 UTC on May 17th, the eye in INSAT-3D had become disorganized. Similarly, on May 18th at 0100 UTC, the eye had also lost its organization in the Microwave satellite. The structural analysis of INSAT-3D can be found in Table \ref{tab2}, and the structural analysis of the Microwave satellite is provided in Table \ref{tab3}.

Two types of eye structures are mainly observed in ESCS, Tauktae: ragged and circular. A large ragged eye, which is non-circular in shape, typically indicates a weak or weakening tropical cyclone. On the other hand, a circular eye can indicate either a weakening cyclone with limited moisture or a weak cyclone overall.

\begin{table}[ht]
    \centering
    \begin{tabular}{|c|c|c|c|}
    \hline
     Date/Time(UTC)&Category & Eye Temp.(degree C) & Eye Structure  \\
     \hline
     16/0600& VSCS & --& First Eye Pattern Seen \\
     16/0900 &VSCS & - 12.8 & Become Ragged \\
     16/1200 & VSCS & - 37.0 & Ragged Eye \& warm \\
     17/0300 & ESCS & - 16.0 & Ragged Eye \\
     17/0600 & ESCS & - 16.0 & No Significant change \\
     17/0900 & ESCS & - 46.0 & Warmest Ragged Eye \\
     17/1500 & ESCS & - 9.0 & Large Ragged Eye \\
     17/1800 & VSCS & - 13.0 & Decrease in size \\
     17/2100 &VSCS & --& Eye is Disorganised \\
     \hline
\end{tabular}
\vspace{2mm}
    \caption{INSAT-3D satellite images used to analyze the structure of the TC eye.}
    \label{tab2}
\end{table}

\begin{table}[ht]
    \centering
    \begin{tabular}{|c|c|c|}
    \hline
       Date/Time(UTC)  & Category & Eye Structure  \\
       \hline
        15/1300 & SCS & First Eye Pattern Seen \\
        16/0800 & ESCS & Ragged Eye \\
        16/1030 & ESCS & Ragged Eye with reduced size \\
        16/1300 & ESCS & Size increase \\
        17/0800 & ESCS & Circular size \\
        17/1100 & ESCS & No Significant Change \\
        17/1300 & ESCS & Circular and size increase \\
        18/0100 & ESCS & Eye is Disorganised \\
        \hline
    \end{tabular}
    \vspace{2mm}
    \caption{Microwave satellite (PWM) images used to analyze the structure of the TC eye.}
    \label{tab3}
\end{table}
\subsubsection{Timeframe for `eye scenes'}
Our analysis of INSAT-3D/3DR images taken every three hours, as well as the available PMW images, revealed frequent sightings of the eye over ESCS, Tauktae once wind speeds reached 60 knots. This observation is consistent with previous research \cite{bib9}, which indicates a sharp increase in the frequency of eye scenes between 50 and 70 knots, followed by another increase after reaching 100 knots. These findings align with the patterns identified in \cite{bib5} as well.

As Tauktae entered its weakening phase, the Maximum Sustained Wind (MSW) decreased to 85 knots starting from 1800 UTC on May 17th, resulting in fewer occurrences of the eye scenes. From May 16th to May 17th (up to landfall), a total of 13 three-hourly eye scenes were observed over ESCS for approximately 39 hours, which is significantly higher than the normal occurrence \cite{bib9}.

According to knapp et al.\cite{bib9}, the average duration of contact between an eye and a typical cyclonic storm is around 30 hours. However, there are storms that exhibit a higher number of eye scenes throughout their lifespan. For instance, the SuCS and AMPHAN storms had 57-hour and 98-hour eye scenes, respectively, as documented in \cite{bib9}.
\subsubsection{In the Eye region, SSTs are occurring}
To assess the relationship between Sea Surface Temperature (SST) and the eye of the storm, as well as Rapid Intensification (RI), SSTs in the Indian Ocean were investigated. During the presence of ESCS, Tauktae, an SST of 30 degrees Celsius was recorded \cite{bib11}. Observations have indicated that SSTs near the circulation of ESCS are consistently above 30.0 degrees Celsius. Similarly, in the Atlantic Ocean, the average SST within the eye is reported as 29.2 degrees Celsius according to \cite{bib6}. In eye scenes observed over the South China Sea, \cite{bib8} demonstrates that most SSTs surpass 29.0 degrees Celsius near the storm's circulation.

In summary, the collective findings of \cite{bib12, bib13, bib14} suggest a consistent association between high SSTs (above 30 degrees Celsius) in the Indian Ocean and the RI of TC, Tauktae. This relationship is attributed to various factors, including low wind shear. The research indicates that the warmth of SSTs plays a primary role in intensifying tropical cyclones by enhancing sensible and latent heat fluxes.
\subsubsection{A characteristic of the eye is its roundness}
Based on the analysis conducted in accordance with \cite{bib15}, the roundness of the eye was determined by evaluating its Equivalent Roundness Value (ERV). Lower ERV values correspond to rounder eyes in less intense tropical cyclones (TCs). It has been observed that the ERV of the eye generally increases with its maximum diameter, indicating that smaller diameter results in a rounder eye \cite{bib8}.

At 2100 UTC on May 16th, when TC intensified into ESCS, an ERV of 0.34 was calculated. The corresponding semi-major and semi-minor axes were measured as 0.08 degrees and 0.085 degrees, respectively. This finding aligns with the conclusion drawn by \cite{bib15} regarding the minimum ERV values.

According to the analysis by \cite{bib15}, hurricanes of Category 5 exhibit higher circularity in their eyes, with an ERV of approximately 0.36. TC, Tauktae falls under Category 4 on the Saffir-Simpson Scale. The mean ERV for TC, Tauktae was found to be 0.57, with dominant values ranging from 0.5 to 0.7. Similar results were observed in the Atlantic Basin, where the predominant ERV values ranged from 0.5 to 0.7, and the mean ERV was also determined as 0.57 \cite{bib15}. Additionally, \cite{bib8} concurs that ERV values of 0.5 to 0.7 are commonly observed in TCs over the South China Sea, which is consistent with the findings for Tauktae.
\section{Final Thoughts} \label{sec4}
The results and discussions lead to the following general conclusions regarding the eye and intensity of TC, Tauktae, along with its growth and evolution, as well as its geometric characteristics.

ESCS, Tauktae exhibited a faster development of the eye, reaching tropical storm intensity within 39 hours, which is shorter than the conventional 48-hour window for eye development. Prior to the formation of the eye, Tauktae displayed a Central Dense Overcast (CDO) pattern at a T4.0 stage.

Interestingly, the eye of ESCS, Tauktae was first observed on microwave imagery approximately 17 hours before becoming visible on INSAT-3D imagery. Real-time monitoring of microwave images allowed us to make predictions based on the lifespan data of Tauktae's intensification.

During the eye development phase, the intensity of ESCS, Tauktae was 45 knots with a Estimated Central Pressure (ECP) of 990 hPa, similar to Atlantic TCs and South China TCs. However, in contrast to the average duration of 30 hours for eye scenes in the Atlantic, ESCS, Tauktae's eye scenes lasted for 39 hours.

The variability of TCs in the North Atlantic and North Pacific oceans aligns with the observations of Eye Roundness Value (ERV). Most ERVs for ESCS, Tauktae ranged from 0.5 to 0.7, with a mean value of 0.57, which is typical for the Indian Ocean region.

\section*{Declarations}
\subsection*{Conflict of Interest}
Conflict of interest is not declared by any of the authors.

\subsection*{Author Contribution}
Research and manuscript preparation were done equally by each author.
\bibliography{sn-bibliography}


\end{document}